\documentstyle[aps,epsf,twocolumn]{revtex}
\def\la{\langle} 
\def\ra{\rangle} 
\def\be{\begin{eqnarray}} 
\def\ee{\end{eqnarray}}

\newcommand{\eq}{\begin{equation}} 
\newcommand{\eqx}{\end{equation}}

\newcommand{\eqn}{\begin{eqnarray}} 
\newcommand{\eqnx}{\end{eqnarray}}
\begin{document}
\draft

\title{\bf Zero Color Magnetization in QCD Matter}

\author{ {\bf Ismail Zahed}$^1$ and {\bf Daniel Zwanziger}$^2$}
\address{$^1$Department of Physics and Astronomy, SUNY, Stony Brook, 
New York 11794, USA.\\ 
$^2$Physics Department, New York University, New York 10003, USA.}
\date{\today} 
\maketitle

\begin{abstract}
We show that all spatial gluon connected correlation functions in 
SU(N) or SO(N) QCD vanish at finite temperature and zero momentum 
in lattice Landau or Coulomb gauges, due to the proximity of the
Gribov horizon. These observations also apply to QCD with two colors 
and an even number of flavors 
at large chemical potential. These nonperturbative
results may have consequences on the nature of the thermal magnetic mass 
and the character of the magnetic color superconductivity.
\end{abstract}

\pacs{}
%PACS numbers : 11.30.Rd, 11.15.Pg, 12.38.-t, 11.30.Qc}

{\bf 1.}
At high temperature or density, the physics of
color charge screening and oscillation in QCD is
closely related to the behaviour of the 
gauge-fixed gluonic correlation functions~\cite{YAFFE}. 
Finite temperature QCD perturbation theory has
proven to be useful but limited to leading
orders due to the persistence of infrared
singularities~\cite{LINDE}. Finite density QCD
perturbation theory is believed to be infrared
tamed due to Landau-damping~\cite{SON}.

Lattice Yang-Mills simulations of the spatial 
Wilson loops show an area law above the critical 
temperature~\cite{POLONYI}, an indication that
magnetic gluons suffer little change across the
deconfining temperature. The electric gluons 
appear to be screened as is evident in the 
behaviour of the Polyakov loop correlation
functions across the critical 
temperature~\cite{ENGELS}.

In this note we show that a lattice regularized
version of QCD at finite temperature, yields zero
for the connected magnetic correlation functions
in the infinite volume limit, both in Landau and
Coulomb gauges, extending an early observation by
Zwanziger~\cite{DAN} at zero temperature. This 
result applies to both SU(N) and SO(N) gauge groups 
irrespective of the center Z$_N$. In short, in Landau 
and Coulomb gauges, the proximity of the Gribov horizon~\cite{GRIBOV}
in the infrared directions puts strong bounds on the 
mean color magnetization~\cite{BEAU}.
Fortunately, no such bounds can be made for the connected 
electric correlation functions, which are expected to
enjoy perturbative screening.

In general, our arguments do not extend to QCD with
finite chemical potential, as the measure on the 
gauge-fixed configurations is no longer positive. An
exception is QCD with two colors and an even number of 
flavors for which 
positivity is recovered. In this case, both the electric and 
magnetic gluon connected correlation functions vanish at zero 
momentum in Landau gauge, 
an indication that nonperturbative physics permeates 
the problem at even large chemical potential.

\vskip .25cm

{\bf 2.} 
Our arguments will follow closely those given
by Zwanziger~\cite{DAN}, and we refer the reader
to them for more details. In short, in the minimal
Landau and Coulomb gauges the fields are selected
by the two conditions~\cite{DAN}:

\be
\bigtriangleup\cdot A = 0\qquad\qquad {\bf K} (A)\geq 0
\label{1}
\ee
where $A$ is the classical lattice connection

\be
A_{\mu} (x) = U_{\mu} (x) -U_{\mu}^{\dagger}(x)
\label{2}
\ee
which is Lie group valued. $\bigtriangleup$ is the lattice divergence.
The $U$'s are SU(N) valued link 
variables from $x$ to $x+e_{\mu}$ on a D-dimensional asymmetric 
lattice $V=\beta\times L^{D-1}$ with periodic boundary conditions.

In Landau gauge, we may follow~\cite{DAN} and use the positivity
of the Faddeev-Popov kernel ${\bf K}$ to bound the zero frequency 
component of the lattice gluon field $A$. Using plane-waves, the 
bounds are~\cite{NOTE1}
\be
|A_4| = &&\left| \frac 1V \sum_x A_4 (x) \right| \leq 2\,{\rm tan} (\pi/\beta)
\nonumber\\
|A_i |= &&\left| \frac 1V \sum_x A_i (x) \right| \leq 2\, {\rm tan} (\pi/L)
\label{3}
\ee
for actually any SU(N) gauge group. The present bounds hold for 
SO(N) as well. While the mean of the spatial color magnetization
vanishes as $L\rightarrow\infty$, its temporal analogue does not at finite 
temperature since $\beta=1/T$ is kept fixed.

Bounds on the free energy follow from the partition function

\be
Z(H) =\int \,d[A]\, \rho (A)\, {\rm exp}\left( \sum_x H\cdot A (x) \right)
\label{4}
\ee
where $A$ is selected by the conditions (\ref{1}), $\rho (A)$
is the QCD measure~\cite{NOTE2}, and
$H^a=(H^a_4,\vec{H}^a)$ a constant colored magnetic field. At finite 
$\beta$ the measure is positive, a requirement for the following
bounds to hold. The free energy $w(H_4,\vec{H})={\rm ln} Z (H)/V$
depends separatly on $H^a_4$ and $\vec{H}^a$. A rerun of the arguments in
\cite{DAN} shows that

\be
&&0\leq w (H_4, 0) \leq 2\,{\rm tan} (\pi/\beta) |H_4|\nonumber\\
&&0\leq w (0,\vec{H}) \leq  2\,{\rm tan} (\pi/L) |\sum_i H_i|
\label{5}
\ee
As $L\rightarrow\infty$, the second inequality implies 
$w(0,\vec{H})=0$. The first inequality remains unaffected. 
The system responds to a constant $H^a_4$ but not to 
a constant $\vec{H}^a$. It follows that all spatial gluon
connected correlation functions vanish at zero three-momentum
in lattice Coulomb gauge, and zero four-momentum in lattice Landau
gauge. In the latter case, the limit is understood in the screening 
sequence as $\omega\rightarrow0$ then $|\vec{k}|\rightarrow 0$, with 
$k=(\omega, \vec{k})$. It is important that the temporal gluon 
correlation functions are not bounded. Indeed, screening of colored
gluons with a finite electric mass takes place in the high temperature
phase~\cite{ENGELS}.

The vanishing of the zero-momentum spatial gluon propagator in Landau
gauge, takes the following form in the infinite volume limit
\be
D_{ii} (0) = \lim_{|k|\rightarrow0}
\lim_{\omega\rightarrow0}
\int d^D x\,\, e^{ik\cdot x}\la A_i (x) \, A_i (0) \ra  = 0
\label{6}
\ee 
in the unrenormalized case. For $D<4$, the renormalization
constants are finite and (\ref{6}) holds in the continuum
limit. For $D=4$, the diverging character of the renormalization
constants prevent us from making similar statements for the
renormalized  propagator in the continuum limit. 
For $|\vec{k}|\sim 0$, (\ref{6}) 
suggests a continuum behaviour analogous to 
$|\vec{k}|^2/(|\vec{k}|^4+ m_M^4)$, that 
is screening masses that are moved to infinity at zero momentum,
in line with Gribov's suggestion in the vacuum~\cite{GRIBOV}.
Given the infrared problem noted in QCD perturbation 
theory~\cite{LINDE}, we expect $m_M\sim g^2 T$.

Recently, lattice simulations of the gluon propagator in
Landau gauge have been carried out on a symmetric lattice
in three-dimensions~\cite{CUCCHI}, and on an asymmetric
lattice in four-dimensions~\cite{HELLER}. The results in
three-dimensions and for small momenta are in agreement 
with the original suggestion by Zwanziger~\cite{DAN}, and
consistent with our results. The fit to the lattice
data in~\cite{HELLER} was forced to $e^{-m_M z}$, while the
analogue of the Gribov propagator at finite temperature
suggests a fit to
\be
\sum_{z_\bot}
D_{ii} (z_\bot ,z) 
\sim e^{-m_M z/\sqrt{2}} \, {\rm cos}\left( \frac {m_M z}{\sqrt{2}} 
+\frac {\pi}4\right)
\label{tale}
\ee
This may explain the z-dependence
observed in the reading of the magnetic mass in~\cite{HELLER}.
The pre-exponent in (\ref{tale}) indicates a change in sign in
the propagator at large distance. Some subtleties related to 
the extrapolation to small momenta on a finite lattice have been 
discussed in~\cite{CUCCHI}.

\vskip .25cm

{\bf 3.}
The extension of the present observations to finite chemical
potential runs into the difficulty that the fermion determinant
is no longer real, making the measure $\rho (A)$ in (\ref{5}) 
complex. However, the case $N_c=2$ and even $N_f$ is an exception.
In general, for $N_c=2$ the fermion determinant admits an extra
symmetry under charge conjugation (assuming that the continuum
Dirac operator is recovered on a fine lattice). If $\lambda$ is an
eigenvalue, so is $-\lambda$ by chiral symmetry, and 
$\lambda^*$ by the invariance under $\tau_2 {\bf C}^{-1} {K}$
where ${K}$ is complex conjugation and ${\bf C}$ is charge
conjugation. Not all eigenstates are four-fold bunched. The
purely real or purely imaginary eigenstates are only 
two-fold bunched. Hence the fermion determinant is real. For even $N_f$ 
it is even positive. In this case, most of the
arguments presented in~\cite{DAN} applies for both the temporal
and spatial correlations of the gluon connected correlation functions
at $\beta=L$ (symmetric lattice). The precedent arguments apply
to the spatial correlation functions for $\beta\neq L$ (asymmetric 
lattice).

In a recent argument~\cite{SON} it was suggested that 
at large chemical potential $\mu$, color magnetic 
superconductivity can be sustained by Landau-damping.
Below the light cone, the spatial gluon 
correlator follows from QCD perturbation theory in the form
$1/(k^2 - im_E^2 \omega/|\vec{k}|)$ with $m_E\sim g\mu$. The
present arguments suggest that below the light cone the 
propagator for the spatial gluons may be analogous to the
one suggested by Gribov, e.g. 
$|\vec{k}|^2/(|\vec{k}|^4+ m_*^4)$, with permanent
`confinement' of magnetic gluons in the infrared. 
We expect $m_*\leq m_E$. For $N_c=2$ and $N_f$ even
perturbation theory  fails below the light cone
at even large chemical potential. Could this affect
the perturbative arguments for QCD with three colors 
at large chemical potential~\cite{SON}?

\vskip .25cm

{\bf 4.}
We have suggested that the proximity of the Gribov
horizon at finite temperature, and also for a special 
case of finite chemical potential, forces a vanishing of
the color magnetization in QCD in both Landau and Coulomb 
gauges. On the lattice, this observation holds for any SU(N) 
or SO(N) gauge group, making the issue of the center $Z_N$ 
not important for this problem. 
In the continuum, this forces the spatial and 
connected gluon correlations to vanish at zero three-momentum, 
suggesting magnetic screening masses that become infinite in the infrared. 
These observations are relevant to the current arguments related 
to the onset of a magnetic mass in QCD at high temperature. 
They also suggest that QCD perturbation theory in the magnetic sector
may not apply even at large chemical potential.

\vskip 1cm

{\bf Acknowledgments}

\vskip 0.25cm

We thank the Aspen School for Theoretical Physics where this work was
started.  This work was supported in part by the US DOE grant
DE-FG02-88ER-40388 and the NSF grant PHY 9520978.


\begin{thebibliography} {99}



\bibitem{YAFFE}
D.J. Gross, R.D. Pisarski and L.G. Yaffe,
Rev. Mod. Phys. {\bf 53} (1981) 43.

\bibitem{LINDE}
A.D. Linde, Phys. Lett. {\bf B96} (1980) 289.

\bibitem{SON}
D.T. Son, hep-ph/9812287.
%R.D. Pisarski  and D.H. Rischke, nucl-th/9811104 (footnote 11).

\bibitem{POLONYI}
E. Manousakis and J. Polonyi, Phys. Rev. Lett. {\bf 58} (1987) 847;
L. Karkkainen et al. Phys. Lett. {\bf B312} (1993) 173.


\bibitem{ENGELS}
J. Engels, V. Mitryushkin and T. Neuhaus, 
Nucl. Phys. {\bf B440} (1995) 555.


\bibitem{DAN}
D. Zwanziger, Phys. Lett. {\bf B257} (1991) 168;
D. Zwanziger, Nucl. Phys. {\bf B364} (1991) 127.

\bibitem{GRIBOV}
V. Gribov, Nucl. Phys. {\bf B139} (1978) 1.

\bibitem{BEAU}
An alternative infrared regularization is 
provided in
L. Baulieu and M. Schaden, Int. Mod. Phys. 
{\bf A13} (1998) 985; M. Schaden Phys. Rev. 
{\bf D58} (1998) 25016.

\bibitem{NOTE1}
A bound can be derived for the finite
momentum components of the gauge-field, from
which the zero momentum limit follows. At finite
temperature, the arguments follow closely those
of Appendix A in~\cite{DAN} (second reference).

\bibitem{NOTE2}
In lattice Coulomb gauge, the measure 
includes an integration over the vertical link variables for a fixed
time slice.


\bibitem{HELLER}
U. Heller, F. Karsch and J. Rank, Phys. Lett. {\bf B355} (1995) 511;
Phys. Rev. {D57} (1998) 1438

\bibitem{CUCCHI}
A. Cucchieri, hep-lat/9902023.


\end{thebibliography}
\end{document}